\documentstyle[aps,prd,twocolumn,epsfig]{revtex}
\newcommand{\be}{\begin{equation}}
\newcommand{\ee}{\end{equation}}
\newcommand{\beq}{\begin{eqnarray}}
\newcommand{\eeq}{\end{eqnarray}}

\begin{document}
\twocolumn[
\begin{center}
{\large\bf\ignorespaces
The Laplacian Gauge  Gluon Propagator in $SU(N_c)$}\\
\bigskip
C.~Alexandrou \\
   {\small\it Department of Physics, University of Cyprus, P.O. Box 20537,
CY-1678 Nicosia, Cyprus} \\ 
\bigskip
Ph.~de~Forcrand \\
{\small\it Institut f\"ur Theoretische Physik, ETH H\"onggerberg,
 CH-8093 Z\"urich, Switzerland \\ and \\
CERN, Theory Division, CH-1211 Geneva 23, Switzerland}\\
\bigskip
E.~Follana \\
 {\small\it Department of Physics, University of Cyprus, P.O. Box 20537,
CY-1678 Nicosia, Cyprus} \\ 
\medskip
{\small\rm (Feb. 15, 2002)} \\
\bigskip
\begin{minipage}{5.5 true in} \small\quad
We examine the gluon propagator in the Laplacian gauge 
in quenched  lattice QCD
as a function of  the number of colours.
We observe a weak dependence on $N_c$ over the whole momentum range.
This implies an almost $N_c$-independent gluon pole mass in units of the string tension.

\medskip

\end{minipage}
\end{center}
\vspace{6mm}
]

\section{Introduction}
Lattice QCD  provides a well suited theoretical framework
for the study of nonperturbative quantities as a function
of the number of colours. Although experimentally inaccessible,
QCD for $N_c=2$ and $N_c=4$ is interesting
from the theoretical point of view. Recently $SU(N_c)$ Yang-Mills theories
were investigated for various $N_c$ by two groups~\cite{Teper,Haris}
with emphasis on  the evaluation of string tensions and glueball
masses and their large $N_c$ limit. 
It is the purpose of the present work to examine
the gluon propagator for different  numbers of colours.
Like in our previous work in $SU(3)$, we evaluate the gluon
propagator in the Laplacian gauge which, unlike the more usual Landau gauge, is free of spurious
Gribov copies on the lattice. Therefore no doubt can be cast on
our results due to the lattice gauge fixing procedure.
Furthermore theoretical arguments have been presented
supporting gauge invariance of the pole mass
 of the transverse gluon propagator~\cite{KKR} and   it is 
therefore an interesting quantity
to investigate as a function of $N_c$
even in a particular gauge.
Studying the  dependence of  the gluon mass on $N_c$
has  theoretical implications for
confinement and chiral symmetry breaking. 
It has been argued
that a non-zero gluon mass is connected to vortex condensation
and to the glueball mass~\cite{Cornwall}.
 Since the $N_c$ dependence of the glueballs has
been investigated, a study of the gluon mass can shed light on  their
relation. In Dyson Schwinger studies the form of the gluon propagator 
affects that of the quark propagator, which in turn provides a determination
of  the quark mass and therefore  is connected to chiral symmetry breaking.

In previous studies~\cite{paperI,paperII} 
we considered the gluon propagator in $SU(3)$, and presented evidence for a pole
which survives the continuum limit. In this work
we present results in $SU(2)$, $SU(3)$
and $SU(4)$ at similar couplings. We
refer the reader to ref.~\cite{paperI} for our notation and the
details of our approach.

\section{Laplacian Gauge fixing}

The Laplacian gauge consists of rotating along a fixed orientation
the local color frame built from the lowest-lying eigenvectors of the covariant Laplacian.
It is well defined in the continuum theory but non-perturbative. On the lattice, 
it has the virtue of
being unambiguous (except for {\em genuine} Gribov copies arising from 
accidental degeneracy of
Laplacian eigenvalues), unlike the lattice Landau gauge.

In the case of $SU(2)$, the Laplacian eigenvalues are twofold degenerate~\cite{Vink}
due to charge conjugation symmetry: if $f$ is an eigenvector so is 
$\sigma_2 f^{\star}$. At each lattice site, one can construct from these two vectors
a $2\times 2$ matrix which is then projected onto $SU(2)$ and defines the required gauge
rotation.

In $SU(3)$ and $SU(4)$, we identify the $N_c-1$ lowest-lying eigenvectors of the Laplacian. 
From these eigenvectors we construct at each lattice site an $SU(N_c)$ matrix by using
a Gram-Schmidt orthogonalization, as described in detail in ref.~\cite{paperI}.

\section{Results}
In order to make the comparison of  our results
in $SU(2)$, $SU(3)$ and $SU(4)$ easier,  we consider $\beta$ values
which yield similar lattice spacings in units of 
the string tension $\sigma$. We take $\beta=2.4,\>\> 6.0 $ and $10.9$ for
$SU(2)$, $SU(3)$ and $SU(4)$ respectively,
which give $a\sqrt{\sigma}=0.2634(16)$~\cite{Teper2}, $0.2265(55)$~\cite{Bali}
and $0.2429(14)$~\cite{Teper} for the three gauge groups.
We can then express our $SU(N_c)$ results in ``physical'' units
using $\sqrt{\sigma} = 440$ MeV. 
The inverse lattice spacing is $1.67(1)$,
$1.94(5)$ and $1.81(1)$~GeV for $SU(2)$, $SU(3)$ and $SU(4)$ respectively.

We first examine $SU(2)$.
For this gauge group, several lattice studies of the gluon propagator in the 
Landau and Coloumb gauges exist~\cite{Langfeld,Cucchieri}.
We present here the 
 first study in the Laplacian gauge where no issues of lattice
Gribov copies arise. It is therefore important to have lattice artifacts under
control.
The volume 
dependence is investigated by considering the 
$SU(2)$ gluon propagator for lattices of size
$4^4$, $8^4$, $16^4$, $16^3\times 64$ and $32^4$. The quantity of interest
is  the transverse part, $D(q^2)$, of the gluon propagator.
This is shown in Fig.~1  in lattice units for the various volumes  
after making a cylindrical cut~\cite{Leinweber} in the momenta which eliminates the leading lattice artifacts
 due to lack of rotational symmetry.
For non-zero momenta, we subtract the longitudinal part of the propagator to
project out its transverse part. For zero momentum this subtraction is not
well defined, and we keep the complete propagator.
We include results from the very asymmetric $16^3\times 64$ lattice.
Even in that case, only small finite-size effects are visible in the infrared 
(momenta less than $\sim 0.5$ GeV in ``physical'' units).

\begin{figure}[h]
\begin{center}
\epsfxsize=8.truecm
\epsfysize=8truecm
\mbox{\epsfbox{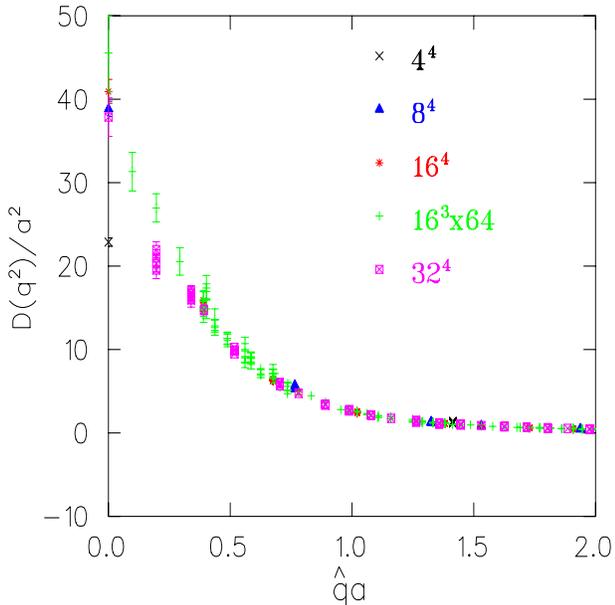}}
\caption{The $SU(2)$ gluon propagator in lattice units for various volumes.}
\end{center}
\label{fig:su2}
\end{figure}

A  similar study  of the 
volume dependence is carried out   for the  SU(4) gluon propagator
for lattices of size
$4^4$, $8^4$, $8^3\times 32$ and  $16^3\times 32$ and the results
are shown in Fig.~2. As for SU(2) we observe only small
  finite-size effects in the infrared
for the very asymmetric $8^3\times 32$ lattice.

\begin{figure}[h]
\begin{center}
\epsfxsize=8.truecm
\epsfysize=8truecm
\mbox{\epsfbox{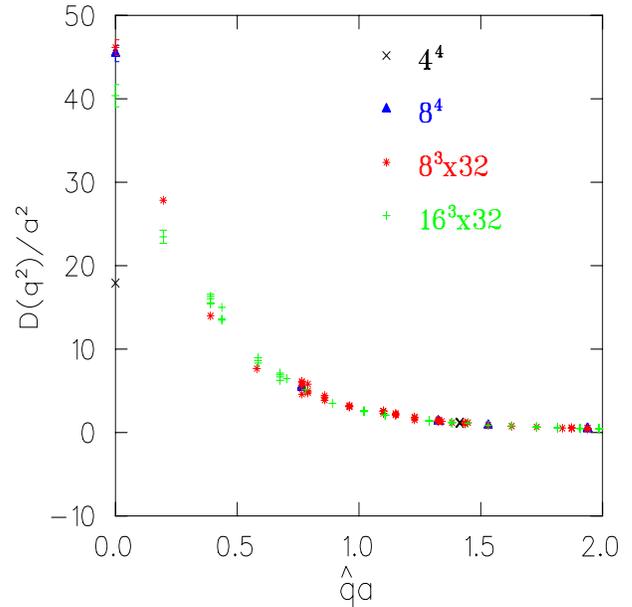}}
\caption{The $SU(4)$ gluon propagator in lattice units for various volumes.}
\end{center}
\label{fig:su4}
\end{figure}

One conclusion to be drawn from Figs.~1 and 2 is
that the zero momentum gluon propagator has a finite value as the
volume is increased. In fact for lattices of spatial size greater than $0.8$~fm
it reaches a constant value.
This behaviour also holds for $SU(3)$~\cite{paperI,paperII}. 
%For $SU(4)$ we have performed simulations
%for only one lattice size  since the similarity of the results in SU(2) 
%and SU(3) leads us to expect the same behaviour also in SU(4).

In Fig.~3 we make a comparison of the renormalized zero momentum gluon propagator,
 ${\cal{D}}(0)$ in ``physical'' units, collecting all our results 
 for the three gauge groups.
To relate the bare lattice propagator to the renormalized continuum propagator
$D_{\rm R}(q;\mu)$
one needs the renormalization constant $Z_3(\mu,a)$:
\be
a^2 D(qa) = Z_3(\mu,a) D_{\rm R} (q;\mu) \quad.
\label{D_R}
\ee
We take as our  renormalization condition 
\be
D_{\rm R}(q)|_{q^2=\mu^2} = \frac{1}{\mu^2}
\label{renormalization point}
\ee
at a renormalization scale $\mu$ which allows a determination of $Z_3(\mu,a)$.
Taking the renormalization 
point at $\mu=1.94$~GeV we find for SU(3)  $Z_3^{SU(3)}(\mu,a)=2.51(1)$,
and for SU(2) and SU(4) the ratios  
$ Z_3^{SU(3)}(\mu,a)/Z_3^{SU(2)}(\mu,a) = 1.07(4)$ and 
 $ Z_3^{SU(3)}(\mu,a)/Z_3^{SU(4)}(\mu,a) = 0.96(2)$ respectively. 

All the 
results for various volumes fall rather well on a universal curve 
demonstrating independence on $N_c$. In particular
both in SU(2) and SU(3) they reach a plateau at approximately similar
volumes.
Since the zero momentum propagator
gives a measure of the range over which the gauge fields are correlated,
these results show that the gauge fields decorrelate in $SU(N_c)$
 at a distance of $ \sim 0.8$~fm 
or about 1.7 times the confining correlation length.

\begin{figure}[h]
\begin{center}
\epsfxsize=7.truecm
\epsfysize=7truecm
\mbox{\epsfbox{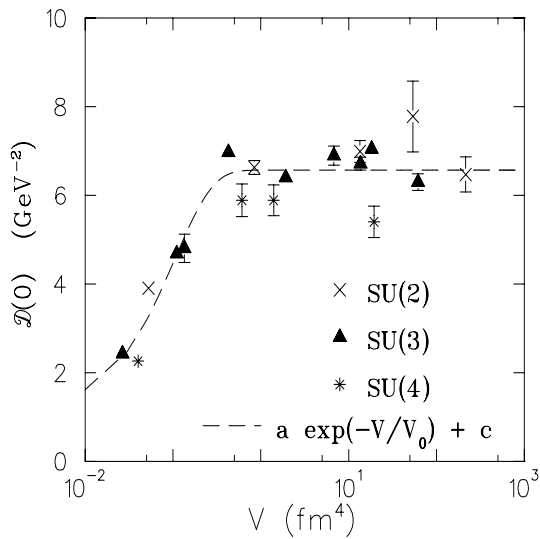}}
\vspace*{0.3cm}
\caption{Volume dependence of the renormalised zero momentum gluon propagator
${\cal{D}}(0)$ for $SU(2)$ (crosses), $SU(3)$ (triangles) and $SU(4)$ (stars).
The dashed line is the best fit to the form $a\exp(-V/V_0)+c$ yielding 
$V_0=0.1$~fm$^4$.}
\end{center}
\label{fig:D0}
\end{figure}

In Fig.~4 we plot the results for the renormalized  transverse 
gluon propagator for lattice sizes for which 
we expect volume effects to be negligible, namely
$16^3\times 32$
for $SU(3)$ and $SU(4)$, and $32^4$ for $SU(2)$.
Remarkably, the overall
shape of the gluon propagator shows no clear $N_c$ dependence.

\begin{figure}[h]
\vspace*{-0.5cm}
\begin{center}
\epsfxsize=7.truecm
\epsfysize=7truecm
\mbox{\epsfbox{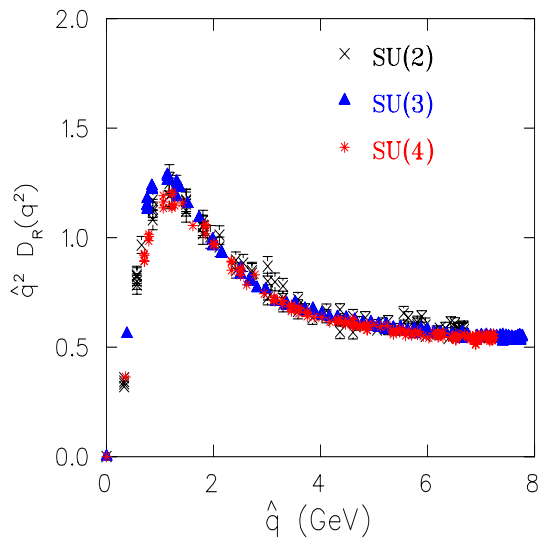}}
\vspace*{0.3cm}
\caption{The momentum dependence of the renormalised transverse
gluon propagator for $SU(2)$ (crosses), $SU(3)$ (triangles) and $SU(4)$ (stars) .}
\end{center}
\label{fig:Nc}
\end{figure}

However, infrared information is suppressed when showing $q^2 D(q^2)$.
To make any low-momentum difference clearly visible,
we plot in Fig.~5 the $SU(2)$ and $SU(4)$ propagators
versus the $SU(3)$ propagator for various momenta. If there would be no 
$N_c$-dependence the results should fall on the diagonal $y=x$. 
Small, but systematic differences are visible between $SU(3)$ and $SU(2) -- SU(4)$ 
where the propagators are the largest, i.e.
in the infrared, for ``physical'' momenta smaller than $\sim 1$~GeV.
The $SU(3)$ propagator is slightly larger, which will translate into a slightly
smaller gluon pole mass.

\begin{figure}[h]
\begin{center}
\epsfxsize=7.truecm
\epsfysize=7truecm
\mbox{\epsfbox{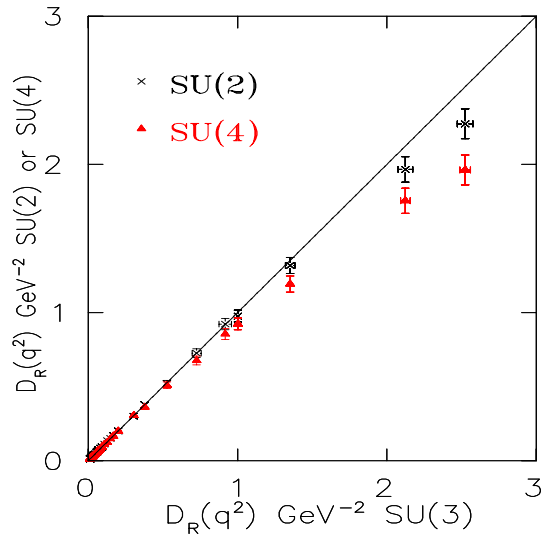}}
\vspace*{0.3cm}
\caption{The renormalised transverse
gluon propagator for $SU(2)$ (crosses) and $SU(4)$ (squares) versus that for
$SU(3)$.}
\end{center}
\label{fig:Nc_comp}
\end{figure}

To investigate more closely the $q$ dependence of the gluon propagator
we perform various fits to phenomenological Ans\"atze. 
We fit the
transverse propagator to Cornwall's Ansatz which allows for a dynamically
generated gluon mass and is
physically motivated. We also fit to two parametrizations
which appeared recently in the literature, namely Model A~\cite{Leinweber}
and  the mass parametrisation of ref.~\cite{Langfeld}: 
\be
D_{\rm{mass}}(q)=\frac{Z}{(q^2+m_1^2)}\Biggl[ \frac{1}{1+q^4/m_2^4} 
   + \frac{s}{\left[log\left(1+q^2/m_L^2\right)\right]^{13/22}}\Biggr]
\ee
In Fig.~6 we 
show the quality of the fits to  the three Ans\"atze: Cornwall's Ansatz has
three fit parameters,  Model A  four and the  mass 
Ansatz five. 
In accord with ref.~\cite{Langfeld} which studied the $SU(2)$ Landau gauge,  
we find that the mass Ansatz works well for $SU(N_c)$. 
This is not so surprising since it has the largest number of parameters.
We find no compelling reason to prefer this Ansatz to the other two.
Cornwall's Ansatz is the only one to allow a 
smooth analytic continuation to $-q^2$ and we will
use it for the determination of the pole mass.

\begin{figure}[h]
\begin{center}
\epsfxsize=7.truecm
\epsfysize=15truecm
\mbox{\epsfbox{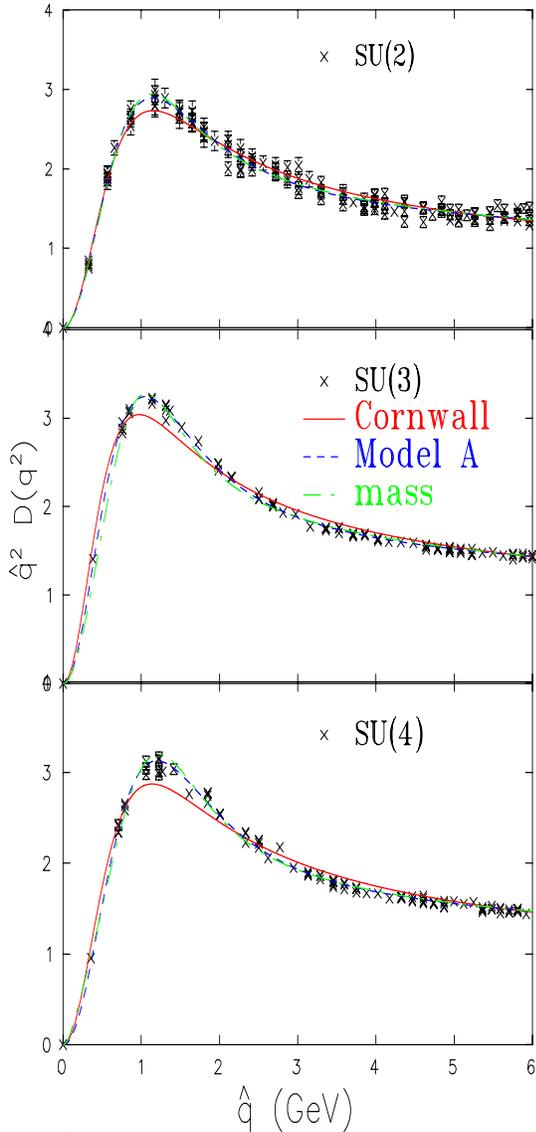}}
\caption{Fits of the gluon propagator to Cornwall's Ansatz (solid line)
, Model A (dashed line) and the mass parametrisation (dash-dotted line), 
in $SU(2)$ (upper), $SU(3)$ (middle) and $SU(4)$ (lower).}
\end{center}
\label{fig:fits}
\end{figure}

The salient features of the infrared behaviour of the 
gluon propagator in the three gauge groups show up by plotting its inverse
as a function of $q^2$.
From Fig.~6 it can be seen that the overall behaviour is very similar
for the three gauge groups although, as shown in Fig.~5,
not identical. To make this observation more quantitative we 
investigate the  gluon pole mass by analytic continuation to negative values of
$q^2$ as shown in Fig.~7. In addition to Cornwall's Ansatz
we use polynomials of increasing degree in  $q^2$ to test the model dependence
of the analytic continuation. 

\begin{figure}[h]
\begin{center}
\epsfxsize=7.truecm
\epsfysize=15truecm
\mbox{\epsfbox{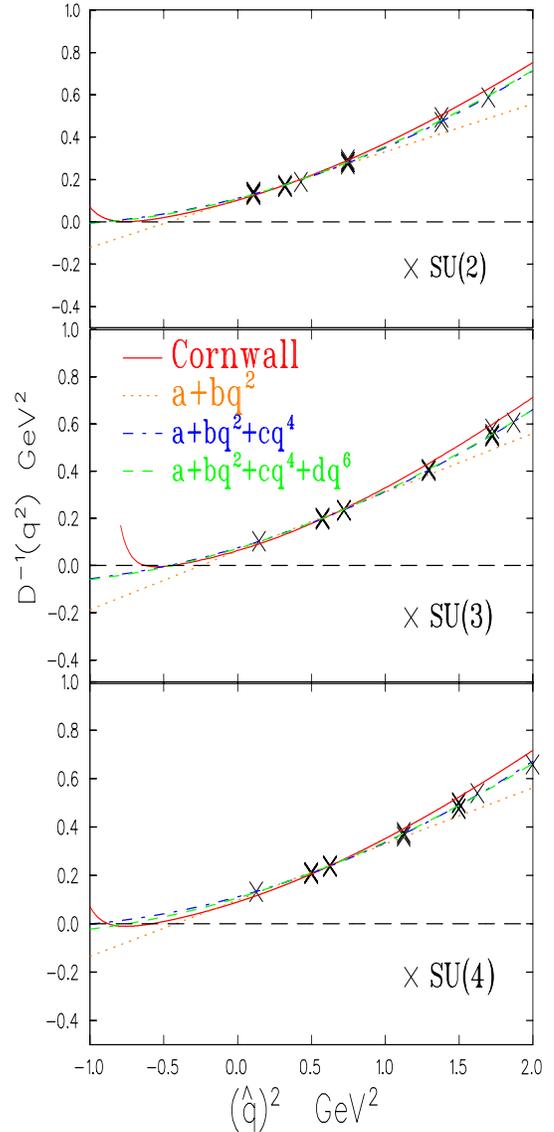}}
\caption{Inverse propagator in the infrared, and analytic continuation to $-q^2$ 
for the determination 
of a pole mass: Upper graph for $SU(2)$, middle for $SU(3)$ and lower for $SU(4)$.
We show Cornwall's Ansatz (solid line), 
a linear (dotted line),
a quadratic (dashed-dotted line) and a cubic (dashed line) polynomial in $q^2$.}
\end{center}
\label{fig:pole}
\end{figure}

Taking as a mean the value extracted from
Cornwall's   Ansatz  and  as
 a crude indication of the systematic error the results of the polynomial fit,
we find for the pole mass 
$(1.85\pm 0.30)\sqrt{\sigma}  $,  $(1.52 \pm 0.06)\sqrt{\sigma} $ 
and  $(1.74 \pm 0.27)\sqrt{\sigma} $ in $SU(2)$,
$SU(3)$ and $SU(4)$ respectively. 
A study of the gluon propagator for $SU(2)$ in Landau gauge~\cite{Langfeld}
found that the low momentum behaviour supported a non-zero
mass of $\sim (1.48 \pm 0.05)\sqrt{\sigma}$.
 Since only the mass parametrisation was used
in that study and no analytic continuation was performed, this
value can be regarded as a rough estimate consistent with our pole mass.

\section{Conclusions}

We have extended a previous lattice study of the gluon propagator in the Laplacian
gauge to $SU(N_c)$. 
In $SU(2)$, we have made an analysis of the finite-size effects, which indicates that
the Laplacian gauge gluon propagator at zero momentum approaches a non-zero constant 
value in the infinite volume limit, as in $SU(3)$.
In fact, this finite non-zero value is independent of $N_c$ within our
statistics. We find ${\cal{D}}(0) = 5.85(5)$~GeV$^{-2}$.

The overall dependence of the gluon propagator on $N_c$ appears extremely weak,
and is barely visible, even for infrared momenta, even for $N_c=2$.
For all $N_c=2,3,4$, the infrared behaviour is consistent with a pole mass in
the range $600-800$~MeV.

Our results give further support to the view 
that $N_c=2$ is a theory already close to the large 
$N_c$ limit of QCD, a conclusion  reached from the study of
the string tension~\cite{Teper} and the glueball mass~\cite{Teper2} in $SU(N_c)$.

\vspace*{0.5cm}

\noindent
\underline{Acknowledgements:}
The $SU(3)$ $16^3\times 32$ lattice configurations were obtained from the
Gauge Connection archive~\cite{connection}.
We thank H. Panagopoulos for discussions.

\end{document}